\documentclass{iopart}
\usepackage{epsf}

\newcommand{\be}{\begin{equation}} 
\newcommand{\ee}{\end{equation}}
\newcommand{\bea}{\begin{eqnarray}} 
\newcommand{\eea}{\end{eqnarray}}

\newcommand{\gton}{\mathrel{\lower.9ex \hbox{$\stackrel{\displaystyle 
>}{\sim}$}}} 
\newcommand{\lton}{\mathrel{\lower.9ex \hbox{$\stackrel{\displaystyle 
<}{\sim}$}}}  

\newcommand{\vp}{{\vec p}}

\begin{document}

\title{Dissipative effects from transport and viscous hydrodynamics}

\author{Denes Molnar$^{1,2}$ and Pasi Huovinen$^1$}

\address{$^1$ Purdue University, Physics Department, 525 Northwestern Ave
West Lafayette, IN 47907, USA}
\address{$^2$ RIKEN BNL Research Center, Brookhaven National Laboratory, 
Upton, NY 11973, USA}

\begin{abstract}
We compare $2\to 2$ covariant transport theory and causal
Israel-Stewart hydrodynamics in 2+1D longitudinally boost invariant
geometry with RHIC-like initial conditions and a conformal $\varepsilon = 3p$
equation of state. The pressure evolution in the center of the
collision zone and the final differential elliptic flow $v_2(p_T)$
from the two theories agree remarkably well for a small shear
viscosity to entropy density ratio $\eta/s \approx 1/(4\pi)$, and
also for a large cross section $\sigma \approx 50$ mb.  A key to this
agreement is keeping {\em all} terms in the Israel-Stewart equations
of motion.  Our results indicate promising prospects for the
applicability of Israel-Stewart dissipative hydrodynamics at RHIC,
provided the shear viscosity of hot and dense quark-gluon matter
is indeed very small for the relevant temperatures $T\sim 200-500$ MeV.
\end{abstract}

\pacs{24.10.Lx, 24.10.Nz, 25.75.Ld}

\section{Introduction}

Recent interest in heavy-ion physics has focused on
constraining the transport properties of hot and dense nuclear matter
using experimental data from RHIC. A particular open question is the
effect of the conjectured ``minimal'' shear viscosity $\eta =
s/(4\pi)$~\cite{AdS/CFT}
on the dynamics and observables ($s$ is the entropy density). 
Studies of dissipation
require a suitable theory framework, in principle a non-equilibrium one.
However, close to local equilibrium, one can also apply
dissipative extensions of ideal (Euler)
hydrodynamics~\cite{romatschke,song,teaney}.

The straightforward Navier-Stokes extension of relativistic ideal
hydrodynamics with corrections linear in gradients leads to acausal
behaviour and instabilities.  An improved formulation proposed by
Mueller and later extended by Israel and Stewart~\cite{IS} (IS)
includes second derivatives, which alleviates the causality
problem. However, that theory originates from an arbitrary truncation
of the entropy current at quadratic order in dissipative corrections
(shear and bulk stress, and heat flow), which is not a controlled
approximation.  Derivations of the IS equations from kinetic theory
again rely on an 
arbitrary truncation of 
nonequilibrium corrections to the phase
space density at quadratic order in momentum (Grad's 14-moment
approximation). In contrast, rigorous (Chapman-Enskog) expansion in
small gradients near local equilibrium results in Navier-Stokes
theory.

Because of these uncertainties about the region of validity of
Israel-Stewart theory, detailed cross-checks against a nonequilibrium
approach are paramount. Here we report on a comparison against
covariant transport theory, for conditions expected in $Au+Au$ at 
$\sqrt{s_{NN}} \sim 200$ GeV at
RHIC, and investigate the effect of small shear viscosities on the
dynamics and differential elliptic flow $v_2(p_T)$.

\section{Covariant transport and Israel-Stewart hydrodynamics}

We solve the equations of motion of causal Israel-Stewart dissipative
hydrodynamics
\bea
\partial_\mu T^{\mu\nu} &=& 0 \ , \qquad \partial_\mu N^{\mu} = 0 \\
D\pi^{\mu\nu} &=& - \frac{1}{\tau_\pi}(\pi^{\mu\nu} - 2\eta
\nabla^{\langle \mu} u^{\nu\rangle}) 
- (u^\mu \pi^{\nu\alpha} + u^\nu \pi^{\alpha\mu}) 
  Du_\alpha  \nonumber  \\
& - &\frac{1}{2} \pi^{\mu\nu} ( \partial_\alpha u^\alpha + 
D \ln \frac{\beta_2}{T} ) 
+ 2 \pi_\lambda ^{\ \langle \mu} \omega^{\nu\rangle\lambda} \label{pieq} \\
T^{\mu\nu} &\equiv & (\varepsilon + p) u^\mu u^\nu - p g^{\mu\nu} + \pi^{\mu\nu} \ ,
\qquad N^\mu \equiv n u^\mu
\eea
in 2+1D longitudinally boost invariant geometry. 
Here $\varepsilon$, $p$, $u^\mu$, and $n$ are the local energy density, 
pressure, flow velocity and particle density; 
$D \equiv u^\mu \partial_\mu$; the $\langle\rangle$ brackets denote
traceless symmetrization and projection orthogonal to the flow
$$
A^{\langle\mu\nu\rangle} 
\equiv 
\frac{1}{2} \Delta^{\mu\alpha}\Delta^{\nu\beta} (A_{\alpha\beta} + A_{\beta\alpha}) 
- \frac{1}{3} \Delta^{\mu\nu} \Delta_{\alpha\beta} A^{\alpha\beta} 
\ , \quad \Delta^{\mu\nu} 
\equiv g^{\mu\nu} - u^\mu u^\nu \ ,
$$
and $\nabla^\mu \equiv \Delta^{\mu\nu}\partial_\nu$.
In the equation for the shear stress $\pi^{\mu\nu}$
we include the term with
the  vorticity tensor
$\omega^{\mu\nu} \equiv (1/2) \Delta^{\mu\alpha}\Delta^{\nu\beta}
(\partial_\beta u_\alpha - \partial_\alpha u_\beta)$
that follows from kinetic theory.

The numerical IS solutions are obtained with a modified version of the
code used in~\cite{pasihydro}.  
Details of the algorithm will be published elsewhere.
Covariant transport solutions are calculated via 
the MPC algorithm~\cite{MPC,v2}.

To aid comparison with covariant
$2\to 2$ transport theory,
we take an ideal gas equation of state of massless particles, 
$\varepsilon = 3p$,
and keep particle number conserved. 
For our conformal 
system bulk viscosity vanishes, while heat flow is ignored for simplicity.
The shear viscosity and shear stress relaxation time of the fluid
are matched to the values from kinetic theory~\cite{IS,DeGroot}:
$\eta \approx  6T /(5\sigma)$ and  
$\tau_\pi \equiv 2 \beta_2 \eta \approx 9\lambda_{MFP}/5$ (because
$\beta_2 = 3/4p$),
where an isotropic two-body cross section is utilized for 
simplicity. We consider two scenarios: i) a constant cross section, and ii) 
a cross
section $\sigma \sim \tau^{2/3}$ growing with proper time 
$\tau\equiv \sqrt{t^2-z^2}$. 
During the initial one-dimensional expansion stage $\eta/s \sim \tau^{2/3}$
grows in the former case, while in the latter case $\eta/s \approx const$.

Initial conditions expected at RHIC for 
$Au+Au$ at $\sqrt{s_{NN}} \sim 200$ GeV and
impact parameter
$b=8$ fm are modelled through an initial density profile proportional
to the (local) number of binary collisions for diffuse gold nuclei,
normalized to $dN/d\eta(b=0) =1000$ to account for the observed
$dN_{ch}/d\eta \approx 700$.  The system is assumed to start from
local thermal equilibrium ($\pi^{\mu\nu} = 0$) at a thermalization time
$\tau_0 = 0.6$ fm with uniform initial temperature
$T_0 = 385$~MeV.

\section{Main results}

\begin{figure}[h]
\leavevmode
\epsfysize=5.5cm
\epsfbox{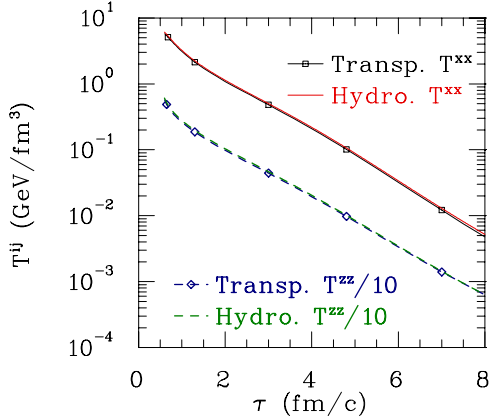}
\caption{Proper time evolution 
of the average transverse ($T^{xx}$, solid) 
and longitudinal pressure ($T^{zz}$, dashed) 
in covariant transport (symbols) and 
causal Israel-Stewart hydrodynamics (no symbols) 
near the center of the collision zone 
($r_T < 1$ fm) for RHIC-like initial conditions (see text) and 
$\eta/s \approx 1/(4\pi$). For clarity, $T^{zz}$ is divided by a factor of 10.
}
\label{Fig:1}
\end{figure}

Figure~\ref{Fig:1} compares the evolution of the transverse and longitudinal 
pressure from covariant transport (lines with symbols) and IS hydrodynamics 
(lines without symbols), averaged over 
the center of the 
collision zone $r_T \equiv \sqrt{x^2+y^2} < 1$ fm for $\eta/s \approx 1/(4\pi)$.
At such a low shear viscosity, we find that
IS hydrodynamics is a good approximation to 
covariant transport in the densest region of the collision.

In order to calculate momentum distributions, a freezeout prescription
is necessary. Here we apply Cooper-Frye (sudden) freezeout
\be
\frac{dN}{dy d^2p_T} = \int d\sigma_\mu(x)  p^\mu f(x,\vp) \ ,
\label{CooperFrye}
\ee
where $f$ is the phasespace density corresponding to the local fluid variables 
and $d\sigma_\mu$ is the local normal to the freezeout hypersurface.
We choose a constant density hypersurface $n = 0.365$~fm$^{-3}$, 
which is the density of an ideal 
gas of massless gluons in chemical equilibrium at $T = 120$ MeV (3 colours).

\begin{figure}[h]
\leavevmode
\epsfysize=5.5cm
\epsfbox{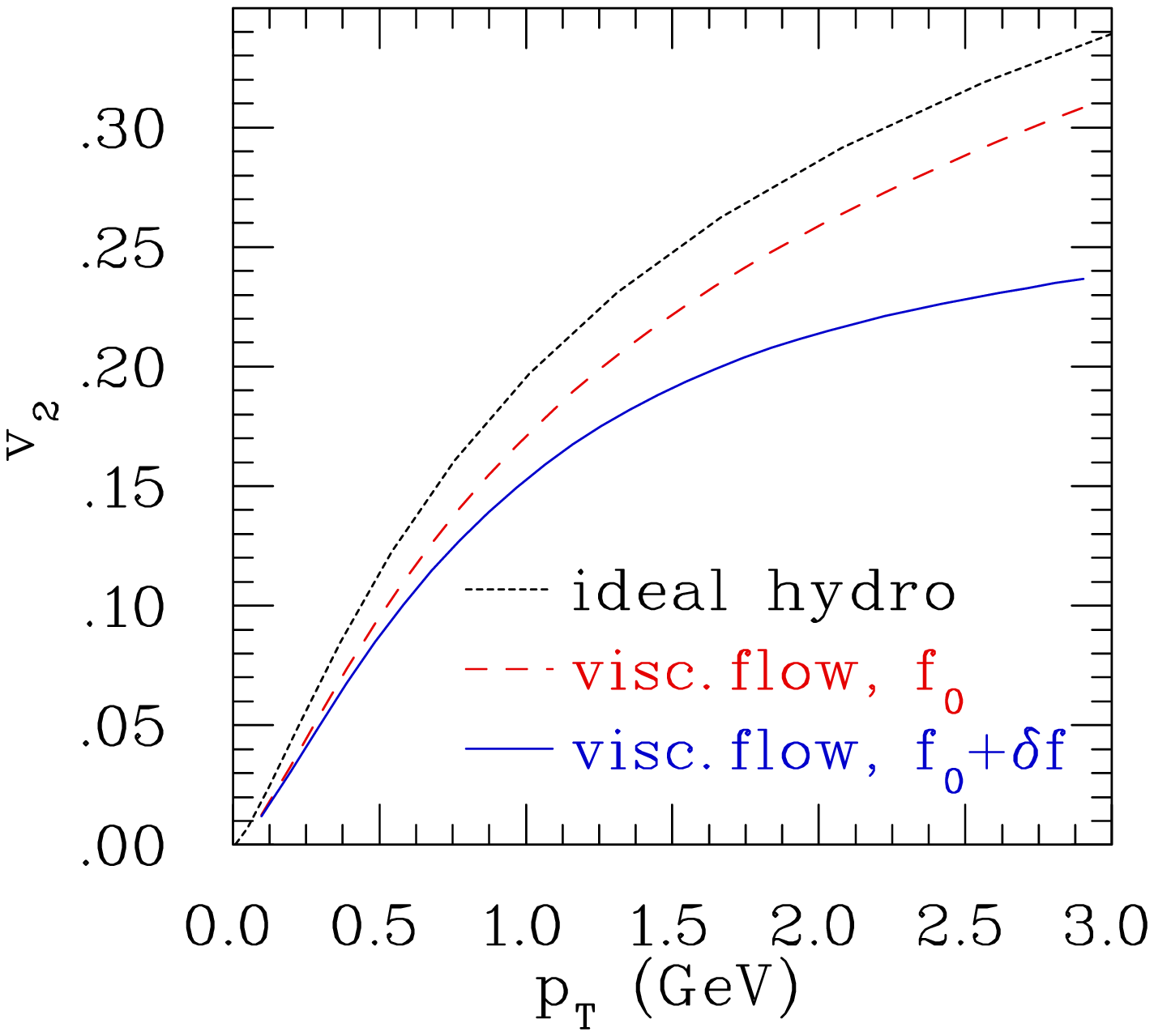}
\epsfysize=5.5cm
\epsfbox{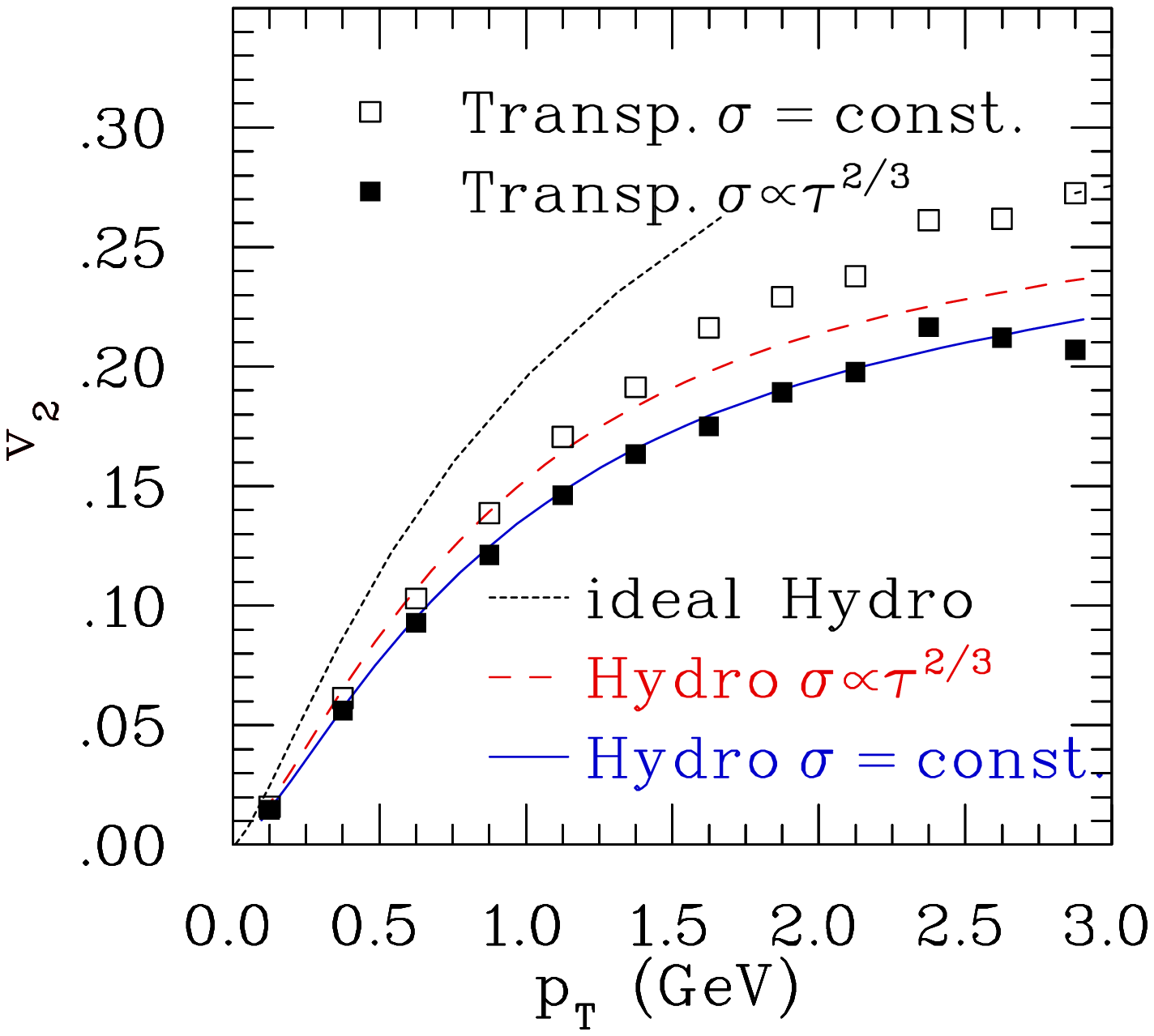}
\caption{
\label{Fig:2}
{\em Left:} Elliptic flow as a function of $p_T$ for 
$Au+Au$ at $\sqrt{s_{NN}} \sim 200$ GeV and
$b= 8$ fm from ideal hydrodynamics (dotted) and IS hydrodynamics 
with Cooper-Frye freezeout ignoring (dashed) or incorporating (solid) 
dissipative
corrections to the local momentum distributions (see text)
for $\eta/s \approx 1/(4\pi)$.
{\em Right:} Comparison of $v_2(p_T)$ from covariant transport (squares) 
and IS hydrodynamics (lines) for $\eta/s \approx 1/(4\pi)$ 
(open squares vs dashed), and $\sigma_{gg\to gg} \approx 47$~mb 
(filled squares vs solid). 
The ideal hydro reference is also shown (dotted).
}
\end{figure}

Figure~\ref{Fig:2}(left) shows elliptic flow from IS hydrodynamics as
a function of transverse momentum, for the $\eta/s \approx 1/(4\pi)$
scenario.  The dashed curve corresponds to (\ref{CooperFrye}) with the
local equilibrium ansatz 
$f_{eq} (x,\vp) = n(x) /[8\pi T(x)^3] \exp[-p^\mu u_\mu/T(x)]$
(we use Boltzmann statistics for consistent comparison with transport).
This incorporates dissipative corrections to the evolution of the flow
field, temperature and particle number density, but ignores
nonequilibrium
distortions of the local momentum distributions (i.e., assumes
$\pi^{\mu\nu} = 0$).  The solid curve is the result for 
\be
f(x,\vp) = f_{eq}(x,\vp)
           \left[ 1+ \frac{p^\mu p^\nu \pi_{\mu\nu}(x)}{8n(x)T^3(x)}\right] 
\ee 
that properly takes the local shear stress into account. We find that
relative to ideal hydrodynamics (dotted curve), dissipation reduces
elliptic flow by $\sim 30$\%, even at such a low
$\eta/s \approx 1/(4\pi)$. This reinforces an earlier kinetic
theory estimate~\cite{minvisc}.  Unlike~\cite{teaney}, we find that
at least one third of the reduction comes from corrections to the ideal
hydrodynamic variables (the lower the $p_T$, the larger the fraction), 
while the remaining up to two-thirds come from shear
stress altering the local momentum distributions.

Finally, we compare in Figure~\ref{Fig:2}(right) 
differential elliptic flow $v_2(p_T)$
between covariant transport (squares) and IS hydrodynamics (dashed and solid).
The dotted line is the ideal hydrodynamics reference.
For 
$\sigma_{gg\to gg} \approx 47$ mb as in~\cite{v2} we find excellent agreement 
between transport and IS hydro (open squares vs dashed line). 
We also find good agreement
for $\eta/s \approx 1/(4\pi)$ (filled squares vs solid line), 
for which IS hydro
somewhat underpredicts the transport results at higher $p_T > 1.5$~GeV. 
We caution that the hydro results at high $p_T$ are also sensitive to the freezeout
prescription, which deserves further investigation.

The modest $\sim$ 30\% dissipative 
corrections found agree well with~\cite{romatschke}
and are much smaller than the effect in~\cite{song}. 
We suspect that this is because~\cite{song}
ignored the first term in the second line of (\ref{pieq}).
If the system is near {\em global} equilibrium, that is appropriate.
On the other hand, RHIC-like initial conditions have large gradients,
and our comparison to kinetic theory shows that in that case 
those terms are important.

\section{Conclusions}

See abstract.

\medskip
\noindent
{\bf Acknowledgements.} We thank RIKEN, 
Brookhaven National Laboratory and
the US Department of Energy [DE-AC02-98CH10886] for providing facilities
essential for the completion of this work.

\end{document}